# Data-Dependent Clock Gating approach for Low Power Sequential System

Dhiraj Sarkar[♭], Pritam Bhattacharjee[†], Alak Majumder[‡]

Integrated Circuit & System (i-CAS) Lab,
Department of Electronics and Communication Engineering
National Institute of Technology, Arunachal Pradesh
Yupia, Dist.– Papumpare, Arunachal Pradesh 791112, INDIA.
[♭]dhirajsarkar026@gmail.com, [†]pritam@ieee.org, [‡]majumder.alak@gmail.com

**ABSTRACT**

Power dissipation in the sequential systems of modern CPU integrated chips (CPU-IC viz., Silicon Chip) is in discussion since the last decade. Researchers have been cultivating many low power design methods to choose the best potential candidate for reducing both static and dynamic power of a chip. Though, clock gating (CG) has been an accepted technique to control dynamic power dissipation, question still loiters on its credibility to handle the static power of the system. Therefore in this paper, we have revisited the popular CG schemes and found out some scope of improvisation to support the simultaneous reduction of static and dynamic power dissipation. Our proposed CG is simulated for 90nm CMOS using Cadence Virtuoso® and has been tested on a conventional Master-Slave Flip-flop at 5GHz clock with a power supply of 1.1Volt. This assignment clearly depicts its supremacy in terms of power and timing metrics in comparison to the implementation of existing CG schemes.

## I. INTRODUCTION

Proper clock distribution inside the sequential systems of CPU (Central Processing Unit) integrated chips (also known nowadays as the Silicon Chip) is very important to setup the uninterrupted functioning of modern computers [1]. The clock handles the periodic signals of 50% duty cycle exemplifying the fact of having highest switching activity [2]. So the power dissipation during the timely propagation of clock signal inside the Silicon Chip is a point of concern as the switching activity factor (α) directly influences the power of the Silicon Chip [3]. In fact, this power dissipation becomes a real threat with the gradual down scaling of CMOS (Complementary-Metal-Oxide-Semiconductor) technologies.

Inside the Silicon Chip, not all the sequential systems or cores start functioning at the same time and hence, the clock supplied to them at that moment is needless. Even more for the case of functionally active systems, the clock signal is required only when there are some significant synchronous operations to be executed. So, the unnecessary clock activity is avoided using a concept called clock gating which is usually implemented in the clocking section of the Silicon Chips. However, the incorporation of clock gating facilitates only the suppression of dynamic power dissipation [4] whereas the controlling of static power dissipation remains unattended. The static power is mainly due to the Off–state transistors in the CMOS circuits of the Silicon Chips [5]. As the clock gating implementation requires lot of transistors, it instigates to a notable amount of static power dissipation. Therefore, a circuital approach is needed which can support the simultaneous reduction of dynamic and static power dissipation. To address the same, we have come up with a new approach of clock gating, which is initially tested on a conventional Master–Slave flip–flop (MS-FF) and later on, the performance is also tested against variation of power supply voltage ($V_{dd}$) and temperature.

Therefore, the framework of the paper is as follows: Section II contains the discussion of the prior works done in clock gating and the motivation to pursue this work. In Section III, the architecture of the proposed clock gating is introduced. The pre and post–layout performance of this proposed gating circuit (implemented on a conventional MS-FF, as it is the popular sequential system inside the modern Silicon Chip) is checked out and a comparative analysis (based on some popular gating schemes) is presented in Section IV. Further, in order to have the knowledge about the versatile functioning of the proposed gating circuit, its performance is checked at different $V_{dd}$ and temperature values. Finally, most of the important observations are concluded in Section V.

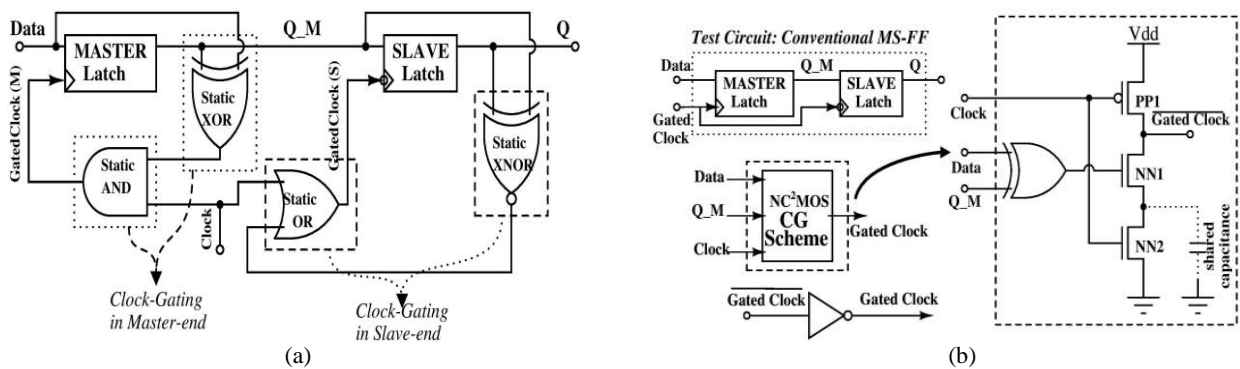

Fig. 1. Popular gating prior–arts: (a) Double–gated clock gating (DG–CG). (b) $NC^2MOS$ clock gating ($NC^2MOS$–CG).

## II. STATE OF THE ART

The concept of clock gating is prevalent in the world of chip designing since the advent of Intel Pentium IV processors [6]. Initially, it was more like the signal dependent clock gating using basic logic gates viz., AND/OR gate [7]. In this case, the gating is done on the basis of an "Enable" signal. Though "Enable" is an independent signal, the role played by it to generate the gated clock has been the major setback for this gating style as the "Enable" signal is always a victim to the environmental noise and can get contaminated with multiple switching [8]. So, the direct use of this is avoided by controlling it using some sequential blocks like latches, flip–flops and registers. This approach is generally referred as the Integrated Clock Gating (ICG) [9]. Even then, the dominance of "Enable" signal in the clock gating process could not be avoided. This real problem commercially brought in the concept of Data–dependent clock gating as a remedy [10] which suggested the triggering of clock only with the logical change in the input data to the sequential block, thereby eliminating the use of "Enable". On the basis of this concept, *Strollo et. al.* had setup their two gating architectures viz., Double–gated clock gating (DG–CG) and negative clocked–CMOS based clock gating ($NC^2MOS$–CG), which were also implemented and tested on conventional Master–Slave based flip–flop (MS–FF) [11] as shown in figure 1(a) and figure 1(b) respectively. Though the $NC^2MOS$ utilizes comparably lesser no. of transistors, it is made of dynamic CMOS principle and hence, the logic strength of the gated clock is mostly improper due to the charge–sharing happening between the intermediate nodes (i.e., NN1 and NN2). So, the $NC^2MOS$–CG is not considered to be an appropriate gating logic. On the other hand in DG–CG, the clock gating is individually carried out for both the Master and Slave latch of MS–FF. This leads to unnecessary gating circuit overhead as the logical data capture happens only at the Master end. A modification of DG–CG has been reported in [12, 13] where the gating overhead is reduced by eliminating the gating components at the Slave end and replacing the static CMOS–based AND gate of Master Latch by a LECTOR–based AND gate. This gating approach (viz., LB–CG) facilitated more reduction of dynamic power along with the control in static power dissipation due to the incorporation of the Leakage Controlled Transistors (LCT) which was reported in [14]. However, it is observed that gating circuit overhead of LB–CG (in terms of the number of transistors) required to construct the gating logic is not impressive. Therefore in this paper, we give a thought to improvise the construction of $NC^2MOS$–CG such that the dynamic state CMOS issue can be avoided.

## III. PROPOSED CLOCK GATING

The schematic of the proposed clock gating is presented in figure 2(a) which is also tested for MS–FF. As per the circuital build-up, the position of the transistor 'NN2' (of figure 1(b)) has now been swapped with the placement of transistor 'N3' (in figure 2(a)). This arrangement of 'N3' facilitates in eliminating the detrimental effects of shared capacitance (as pointed out in figure 1(b)) of NC²MOS–CG. It is found that as the source terminal of n-channel MOS 'N3' is closely connected to the power line (i.e., ground pin of the circuit); the logic '0' at the node $\overline{Gated\_Clock}$ is strong enough to remain within the noise margin. More importantly, the ON and OFF state of 'N3' is controlled by the 'Comp' signal which is generated 'high' only at the time of logical change in the input 'Data'. The 'Clock' cannot approach the 'Gated Clock' when the 'Comp' is low. In the meantime even if 'Clock' switches, it also does not allow the "Contention Current" [15] to flow between the power supply '$V_{dd}$' and ground line through the transistors P1 and N1. Therefore, the proposed CG scheme claims to restrict the static power dissipation to certain extent.

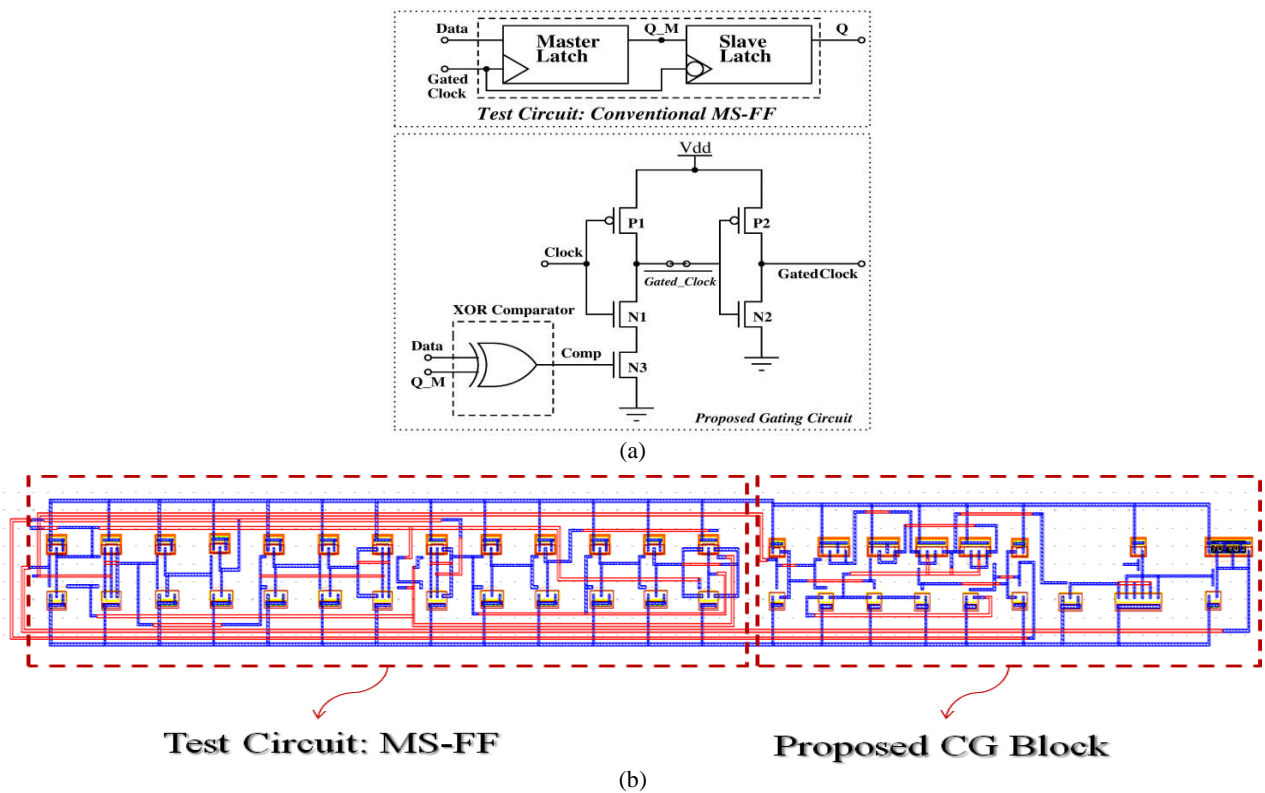

Fig. 2. (a) Schematic and (b) Layout of the Proposed Gating implementation using 90nm CMOS.

TABLE I
PARAMETRIC VALUES USED TO OBTAIN THE TRANSIENT RESPONSE OF PROPOSED CLOCK GATING

| $V_{dd}$ (volts) | Clock | | | Random Data | | | Device Widths (μm) of P1, P2, N1, N2, N3 respectively |
|---|---|---|---|---|---|---|---|
| | Rise time (ps) | Frequency (GHz) | Fall time (ps) | Rise time (ps) | Frequency (GHz) | Fall time (ps) | |
| 1.1 | 20 | 5 | 20 | 20 | 0.206 | 20 | 0.2, 1, 1, 0.2, 0.4 |

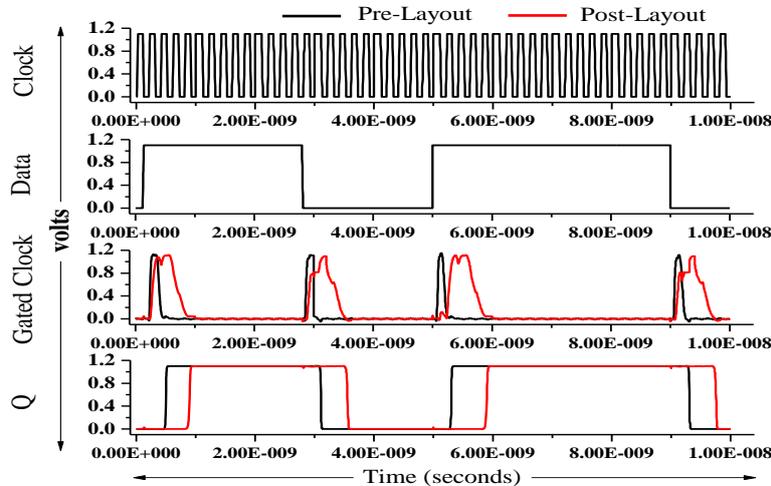

Fig. 3. Transient response of Proposed Gating Circuit using 90nm CMOS.

Here in this paper, the proposed gating circuit is designed in Cadence Virtuoso® using the environment of 90nm process library viz., "GPDK090" [16]. The physical layout of the whole circuital implementation is presented in figure 2(b) which has consumed an area of about 629.23μm², out of which the gating circuit overhead and MS-FF read an area of 233.42μm² and 395.80μm² respectively. In Table I, the parametric values of the circuit construction are presented which has been utilized to obtain the transient response for both the pre and post layout version (depicted in figure 3) of the design. It is evident through the transient response that the generated gated clock successfully synchronizes the 'Data' signal transmission to the output 'Q' of MS–FF and thereby saves the unnecessary clock switching inside the Silicon Chip leading to lesser power dissipation.

## IV. ANALYSIS OF PERFORMANCE METRICS

This section is dedicated for the discussion on circuit performance (for both the pre and post–layout version) of our proposed CG scheme in context to the performances of NC²MOS–CG, LB–CG and No–gating design styles. However, the comparison with DG–CG is not stated here as it is an inferior CG scheme in compare to the above mentioned design styles which is already reported in [12]. Thereby, the discussion is categorized under three individual entities viz., Power Analysis, Timing Analysis and Circuit performance against the variation in $V_{dd}$ and temperature.

### A. *Power Analysis*

The pre-layout version of the proposed gating scheme is found to dissipate an average power of 18.249μW whereas the exerted dynamic and static power is about 1.2367μW/GHz and 12.0655μW respectively. However, the values of average, dynamic and static power in the pre-layout version is individually 59.41%, 77.51% and 30.93% lesser in par to these values obtained against its post-layout version. This difference in the parametric values of the post-layout version can only be controlled with the incorporation of sophisticated layout design techniques [17].

TABLE II
POWER PERFORMANCE OF THE CG SCHEMES @ 5 GHZ CLOCK

| Different CG Schemes | | Average Power (μW) | Dynamic Power (μW/GHz) | Static Power (μW) |
|---|---|---|---|---|
| No–Gating (Pre–Layout) | | 34.265 | 3.2181 | 18.1745 |
| LB–CG (Pre–Layout) | | 22.528 | 1.57027 | 14.6766 |
| NC²MOS–CG (Pre–Layout) | | 23.463 | 1.6521 | 15.2025 |
| Proposed–CG | Pre–Layout | 18.249 | 1.2367 | 12.0655 |
| | Post–Layout | 44.968 | 5.4995 | 17.4703 |

Nevertheless, the individual power values in the pre-layout version of the proposed gating circuit are found to be the improved one in comparison to the values depicted by LB–CG, NC²MOS–CG and the No–gating counterpart (as shown in Table II). For example, the rate of improvement in the values of average, dynamic and static power dissipation is individually 22.22%, 25.14% and 20.63% lesser in comparison to NC$^2$MOS–CG.

### B. *Timing Analysis*

The value of timing parameters (viz., Delay, Setup time and Hold time) are estimated in accordance to the assessment pattern reported in [18]. It is found that propagation delay in the pre-layout version of proposed CG is 7.16% higher than the No–gating counterpart (as given in Table III) but fortunately 9.02% and 1.16% lesser than the delay in LB–CG and NC$^2$MOS–CG respectively. However, the propagation delay inside the post-layout version of proposed CG has comparatively increased by 50.07% maybe due to the inclusion of parasitic elements during the post-layout verification.

TABLE III
TIMING PERFORMANCE OF THE CG SCHEMES @ 5 GHZ CLOCK

| Different CG Schemes | | Delay (ps) | Setup Time (ps) | Hold Time (ps) | Latency (ps) |
|---|---|---|---|---|---|
| No–Gating (Pre–Layout) | | 94.37 | 133.68 | -87.9 | 228.05 |
| LB–CG (Pre–Layout) | | 111.16 | 104.56 | -77.903 | 189.07 |
| NC$^2$MOS–CG (Pre–Layout) | | 102.32 | 84.64 | -79.138 | 186.96 |
| Proposed–CG | Pre–Layout | 101.13 | 117.44 | -81.04 | 218.58 |
| | Post–Layout | 202.58 | 343.052 | -109.51 | 545.63 |

The fact is that the severity of timing criticality in a circuit is well understood with the help of 'Latency' (considered to be the summation of delay and setup time which depict that the increase in the amount of latency, increases the timing criticality [19]). It is observed that the latency of the proposed CG is 4.15% lesser than the No–gating counterpart but 15.60% and 16.91% greater than LB–CG and NC$^2$MOS–CG respectively.

TABLE IV
POWER-DELAY-PRODUCT OF THE CG SCHEMES @ 5 GHZ CLOCK

| Different CG Schemes | | Power-Delay-Product (fJ) |
|---|---|---|
| No–Gating (Pre–Layout) | | 32.33 |
| LB–CG (Pre–Layout) | | 25.04 |
| NC$^2$MOS–CG (Pre–Layout) | | 24.00 |
| Proposed–CG | Pre–Layout | 18.45 |
| | Post–Layout | 91.09 |

However, the power–delay–product (PDP) of the pre-layout version of the proposed CG (shown in Table IV) is 23.12%, 26.31%, 42.93% lesser in comparison to the individual PDP values for NC$^2$MOS–CG, LB–CG and No–Gating counterpart. This result clearly depict that the proposed CG scheme is a suitable candidate to design for low power sequential systems inside the Silicon Chip.

### C. *Circuit performance against varied power supply voltage and temperature*

Nevertheless, the reliability of the proposed CG is needed to be cross-checked which is why the circuit performance is also observed under the variation of $V_{dd}$ and temperature. Firstly, the average power dissipation of proposed CG is seen with the change in $V_{dd}$ as shown in figure 4(a). It is observed that the average power dissipation of the proposed CG is the least among the obtained values of average power in all the other CG schemes. However, upon checking the delay of our proposed CG for varied $V_{dd}$ (as given in figure 4(b)), we have not observed a significant advantage in contrast to the delay of the prior schemes. Therefore, we look into the PDP markings (given in figure 4(c)) of these CG schemes where it is noted that the performance of the proposed CG is the best in accordance to the variation of $V_{dd}$. However on a second note, in figure 5(a), 5(b) and 5(c), the average power dissipation, delay and PDP of all CG schemes are individually observed for varied operating temperature. Although, the implementation of proposed CG scheme in MS–FF looks appropriate in context to the amount of power dissipation, the computational speed (i.e., delay) seems to be a point of concern with the increase in temperature. Therefore, we examine the PDP variation against temperature sweep and observe that the proposed CG scheme has depicted the minimum PDP amongst the implementation of all other CG schemes, inferring it to be the optimal approach to design for low power sequential systems.

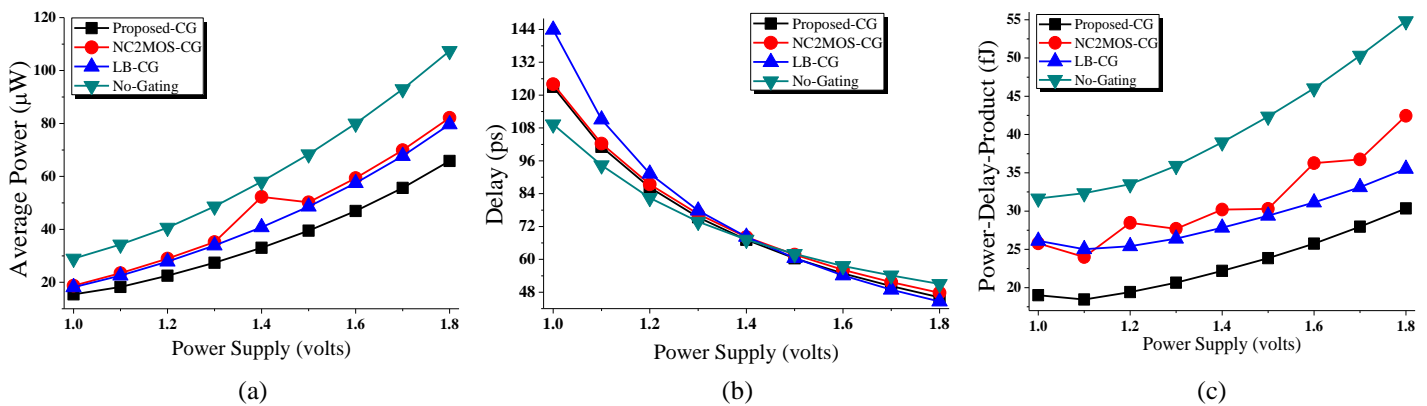

Fig. 4. Variation in (a) Average Power, (b) Delay and (c) PDP @$V_{dd}$.

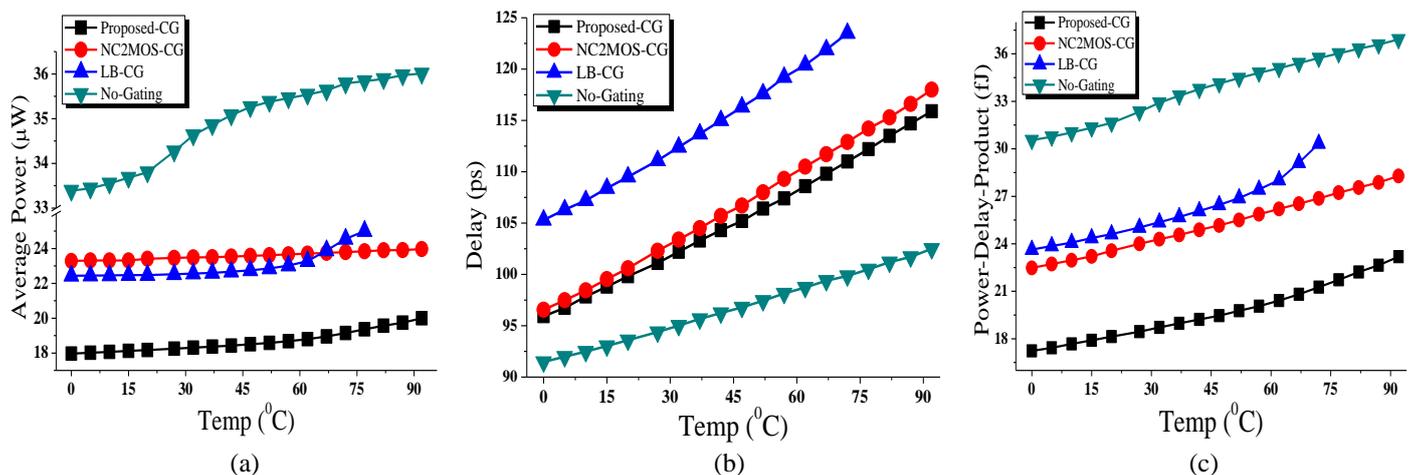

Fig. 5. Variation in (a) Average Power, (b) Delay and (c) PDP @Temp.

## V. CONCLUSION

In this paper, a new architecture of clock gating is presented which is capable of delivering the notable reduction in the average, static and dynamic power dissipation of the conventional master–slave flip flop (known to be the popular sequential element inside the Silicon Chip). The pre–layout version of the proposed CG scheme has 22.22%, 19.22% and 46.74%% average power reduction in comparison to the average power dissipated in $NC^2MOS$–CG, LB–CG and no–gating peer respectively. The rate of improvement of the proposed CG for static and dynamic power dissipation individually is 20.63% and 25.14% (against $NC^2MOS$–CG), 17.79% and 13.52% (against LB–CG) and 33.61% and 61.57% (against the no–gated design). Interestingly, the circuit performance of the proposed CG scheme is also aware of variation in $V_{dd}$ and temperature, depicting it to be the smartest among the existing CG schemes. Therefore, the proposed CG scheme is asserted to be the suitable approach to design for low power sequential applications.

## ACKNOWLEDGEMENT

This work is carried out in support of MEITY, Govt. of India under their Visvesvaraya PhD Scheme & SMDP–C2SD project.